\documentclass{mem}
\usepackage{natbib}
\usepackage{txfonts}
\usepackage{balance}
\usepackage{graphicx}
\usepackage[a4paper]{hyperref}
\idline{80}{1}
\begin{document}

\newcommand{\Hy}{\ensuremath{\null^1{\rm H}}}
\newcommand{\De}{\ensuremath{{\rm D}}}
\newcommand{\D}{\ensuremath{\null^2{\rm H}}}
\newcommand{\Tr}{\ensuremath{{\rm T}}}
\newcommand{\T}{\ensuremath{\null^3{\rm H}}}
\newcommand{\Het}{\ensuremath{\null^3{\rm He}}}
\newcommand{\He}{\ensuremath{\null^4{\rm He}}}
\newcommand{\Be}{\ensuremath{\null^7{\rm Be}}}
\newcommand{\Lis}{\ensuremath{\null^6{\rm Li}}}
\newcommand{\Li}{\ensuremath{\null^7{\rm Li}}}

\title{ WMAP 5-year constraints on time variation of $\alpha$ and $m_e$ }

\author{Claudia G. Sc\'{o}ccola \inst{1,2}, Susana. J. Landau \inst{3}, Hector Vucetich \inst{1} }

\institute{
Facultad de Ciencias Astron\'{o}micas y Geof\'{\i}sicas. Universidad Nacional de La Plata,
Paseo del Bosque S/N 1900,
 La Plata, Argentina.
\and
Max-Planck-Institut f\"ur Astrophysik, Karl-Schwarzschild-Str. 1, 85741 Garching, Germany.
\email{scoccola@MPA-Garching.MPG.DE}
\and
Departamento de F{\'\i}sica, FCEyN, Universidad de Buenos Aires, 
Ciudad Universitaria - Pab. 1,
 1428 Buenos Aires,Argentina
\email{slandau@df.uba.ar}
}

\authorrunning{Sc\'occola et al.}

\titlerunning{WMAP 5 constraints on $\alpha$ and $m_e$} 

\abstract{We studied the role of fundamental constants in an updated
  recombination scenario, focusing on the time variation of the fine
  structure constant $\alpha$ and the electron mass $m_e$ in the early
  Universe. Using CMB data including WMAP 5-yr release, and the 2dFGRS
  power spectrum, we put bounds on variations of these constants, when
  both constants are allowed to vary, and in the case that only one of
  them is variable. In particular, we have found that $-0.019 < \Delta
  \alpha / \alpha_0 < 0.017$ (95\% c.l.), in our joint estimation of
  $\alpha$ and cosmological parameters.  Finally, we analyze how the
  constraints depend on the recombination scenario.}

\maketitle

\section{Introduction}
\label{Intro}

Time variation of fundamental constants is a prediction of theories
that attempt to unify the four interactions in nature. Many
observational and experimental efforts have been made to put
constraints on such variations. Cosmic microwave background radiation
(CMB) is one of the most powerful tools to study the early universe
and in particular, to put bounds on possible variations of the
fundamental constants between early times and the present.

Previous analysis of CMB data (earlier than the WMAP five-year
release) including a possible variation of $\alpha$ have been
performed by \citet{Martins02,Rocha03,ichi06,Stefanescu07,Mosquera07,landau08}
and including a possible variation of $m_e$ have been performed by
\citet{ichi06,Scoccola07,landau08,YS03}.

In the last years, the process of recombination has been revised in
great detail \citep{chluba1,chluba2,Hirata09}, and in particular, helium
recombination has been calculated very precisely, revealing the
importance of considering new physical processes in the calculation of
the recombination history \citep{Dubrovich05,SH08a,SH08b,SH08c,KIV07}.

In a previous paper \citep{Scoccola08}, we have analized the variation
of $\alpha$ and $m_e$ in an improved recombination scenario. Moreover,
we have put bounds on the possible variation of these constants using
CMB data and the power spectrum of the 2dFGRS.  In Section
\ref{dependencies}, we review the dependences on $\alpha$ and $m_e$ in
the recombination scenario.  In section \ref{resultados} we present
bounds on the possible variation of $\alpha$ and $m_e$ using the 5-yr
data release of WMAP \citep{Hinshaw09} together with other CMB
experiments and the power spectrum of the 2dFGRS \citep{2dF05}. In
addition to the results of our previous work \citep{Scoccola08}, here
we also present bounds considering only variation of one fundamental
constant ($\alpha$ or $m_e$) alone.  A comparison with similar
analyses by other authors is presented in Section \ref{conclusion}.

\section{The detailed recombination scenario }

\label{dependencies}

The equations to solve the detailed recombination scenario can be
found for example in \citet{wong08}, as they are coded in R{\sc
  ecfast}. In the equation for helium recombination, a term which
accounts for the semi-forbidden transition $2^3$p--$1^1$s is
added. Furthermore, the continuum opacity of HI is taken into account
by a modification in the escape probability of the photons that excite
helium atoms.

The cosmological redshifting of a transition line photon $K$ and the
Sobolev escape probability $p_{\rm S}$ are related through the
following equation~(taking He\,{\sc i} as an example):
 {\setlength\arraycolsep{1pt}
\begin{eqnarray}
&& K_{\rm HeI} = \frac{g_{{\rm HeI}, 1^1{\rm s}}}{g_{{\rm HeI}, 2^1{\rm p}}}
\frac{1}{ n_{{\rm HeI}, 1^1{\rm s}}  
A^{\rm HeI}_{ 2^1{\rm p}-1^1{\rm s}} p_{\rm S}} 
\label{eqKHeI}
\end{eqnarray}}
 where $A_{{\rm HeI}, 2^1{\rm p}-1^1{\rm s}}$ is the $A$ Einstein
coefficient of the He\,{\sc I} $2^1$p--$1^1$s transition.  To include
the effect of the continuum opacity due to HI, based on the approximate
formula suggested by \citet{KIV07}, $p_{\rm S}$ is replaced by the new
escape probability $p_{esc}=p_s + p_{\rm con, H}$ with
\begin{equation}
p_{\rm con, H} = \frac{1}{1 + a_{\rm He} \gamma^{b_{\rm He}}},
\end{equation}
and
\begin{equation}
 \gamma = \frac{\frac{g_{{\rm HeI}, 1^1{\rm s}}}{g_{{\rm HeI}, 2^1{\rm p}}}
A^{\rm HeI}_{2^1{\rm p}-1^1{\rm s}} (f_{\rm He} - x_{\rm HeII})c^2}
{8 \pi^{3/2} \sigma_{{\rm H},1{\rm s}}(\nu_{\rm HeI,2^1 p}) 
\nu_{\rm HeI,2^1{\rm p}}^2 \Delta \nu_{\rm D,2^1p} 
(1 - x_{\rm p})}\, 
\end{equation}
where $\sigma_{{\rm H},1s}(\nu_{\rm HeI,2^1p})$ is the H ionization
cross-section at frequency $\nu_{\rm HeI,2^1p}$ and $\Delta \nu_{\rm
D,2^1p} = \nu_{\rm HeI,2^1p} \sqrt{2 k_{\rm B} T_{\rm M}/m_{\rm He}
c^2}$ is the thermal width of the He\,{\sc i} $2^1$p--$1^1$s line.

The transition probability rates $A_{{\rm HeI}, 2^1{\rm p}-1^1{\rm
s}}$ and $A_{{\rm HeI}, 2^3{\rm p}-1^1{\rm s}}$ can be expressed as
follows \citep{DM07}:

\begin{equation}
 A^{\rm HeI}_{i-j} = \frac{4 \alpha}{3 c^2} \omega_{ij}^3 \left|\left<\psi_i|r_1 + r_2|\psi_j\right>\right|^2
\label{rate}
\end{equation}
where $\omega_{ij}$ is the frequency of the transition, and $i$($j$)
refers to the initial (final) state of the atom.  It can be shown
\citep{Scoccola08} that to first order in perturbation theory, the
dependence of the bracket goes as the Bohr radius $a_0$.  On the other
hand, $\omega_{ij}$ is proportional to the difference of energy levels
and thus its dependence on the fundamental constants is $\omega_{ij}
\simeq m_e \alpha^2$. Consequently, the dependence of the transition
probabilities of ${\rm HeI}$ on $\alpha$ and $m_e$ can be expressed as
\begin{equation}
A^{\rm HeI}_{i-j} \simeq m_e \alpha^5.
\end{equation}
 The dependences on $\alpha$ and $m_e$ of all the physical quantities
 relevant at recombination are summarized in
 Table~\ref{resumen_depend}.

\begin{table*}
\begin{center}
\renewcommand{\arraystretch}{1.3}
\begin{tabular}{|l|l|l|}
\hline
Description & Physical Quantity  & Dependence   \\
\hline
Binding Energy of Hydrogen  & $B_1$ & {\ \ \ }$\alpha^2 m_e$ \\
\hline
Transition frequencies &  $\nu_{\rm H2s}$,$\nu_{{\rm HeI}, 2^1{\rm s}}$,$\nu_{{\rm HeI},2^3{\rm s}}$ & {\ \ \ }$\alpha^2  m_e$ \\
\hline
Photoionization cross section n &  $\sigma_n(Z,h\nu)$   & {\ \ \ }$\alpha^{-1} m_e^{-2} $  \\
\hline 
Thomson scattering cross section &  $\sigma_T$   & {\ \ \ }$\alpha^2 m_e^{-2} $  \\
\hline 
Recombination Coefficients Case B & $\alpha_{\rm H}$, $\alpha_{\rm HeI}$, $\alpha^t_{\rm HeI}$ &  {\ \ \ }$\alpha^3 m_e^{-3/2}$ \\
\hline 
Ionization Coefficients & $\beta_{\rm H}$, $\beta_{\rm HeI}$ &{\ \ \ }$\alpha^3$
\\ \hline
Cosmological redshift of photons & $K_{\rm H}$, $K_{\rm HeI}$, $K^t_{\rm HeI}$ & {\ \ \ }$\alpha^{-6} m_e^{-3}$ \\
\hline
Einstein $A$ Coefficients & $A^{\rm HeI}_{i-j}$ & {\ \ \ }$\alpha^5 m_e$ \\
\hline
Decay rate 2s $\rightarrow$ 1s  & $\Lambda_{\rm H}$, $\Lambda_{\rm HeI}$ & {\ \ \ }$\alpha^8 m_e$ \\
\hline
\end{tabular}
\caption{Dependence on  $\alpha$ and $m_e$ of the physical quantities relevant during recombination.}
 \label{resumen_depend}
\end{center}
\end{table*}

\begin{figure*}[t!]
\resizebox{\hsize}{!}{
\includegraphics[scale=0.01,angle=-90]{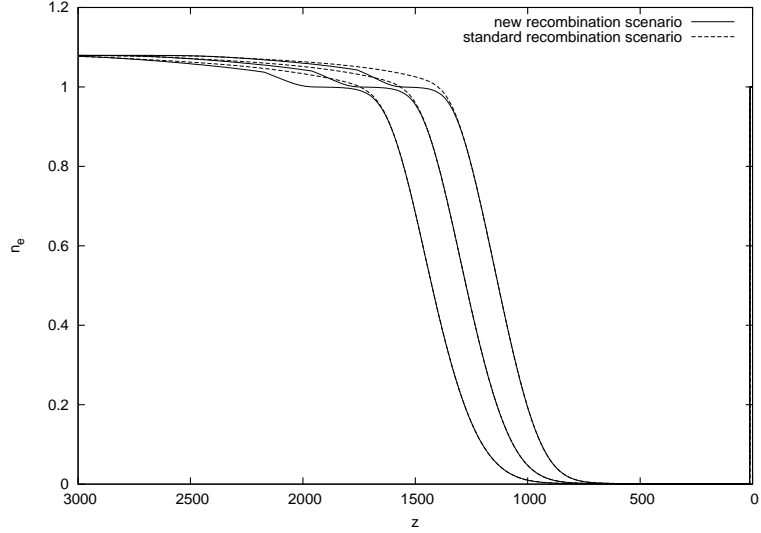}}
\caption{Ionization history allowing $\alpha$ to vary with time. From left to right, the values of $\frac{\alpha}{\alpha_0}$ are 1.05, 1.00, and 0.95, respectively. The dotted lines correspond to the standard recombination scenario, and the solid lines correspond to the updated one.}
\label{ionization_history}
\end{figure*}

Fig.~\ref{ionization_history} shows the ionization history for
different values of $\alpha$. Recombination occurs at higher redshift
if $\alpha$ is larger.  On the other hand, there is little change when
considering different recombination scenarios, for a given value of
$\alpha$.  Something similar happens when varying $m_e$.

With regards to the fitting parameters $a_{\rm He}$ and $b_{\rm He}$,
it is not possible to determine yet the effect that a variation of
$\alpha$ or $m_e$ would have on them. However, we have shown in
\citet{Scoccola08} that for the precision of WMAP data, there is no
need to know these dependences.

\section{Results}
\label{resultados}

We performed our statistical analysis by exploring the parameter space
with Monte Carlo Markov chains generated with the CosmoMC code
\citep{LB02} which uses the Boltzmann code CAMB \citep{LCL00} and
R{\sc ecfast} to compute the CMB power spectra. We modified them in
order to include the possible variation of $\alpha$ and $m_e$ at
recombination. 
 Results are shown in Table \ref{tablacmb}.

We use data from the WMAP 5-year temperature and
temperature-polarization power spectrum \citep{Nolta08}, and other CMB
experiments such as CBI \citep{CBI04}, ACBAR \citep{ACBAR02}, and
BOOMERANG \citep{BOOM05_polar,BOOM05_temp}, and the power spectrum of
the 2dFGRS \citep{2dF05}. We have considered a spatially-flat
cosmological model with adiabatic density fluctuations, and the
following parameters:
\begin{equation}
P=\left(\Omega_B h^2, \Omega_{CDM} h^2, \Theta, \tau, \frac{\Delta
\alpha}{\alpha_0}, \frac{\Delta m_e}{(m_e)_0}, n_s, A_s\right) \nonumber \, \, 
\end{equation}
where $\Omega_{CDM} h^2$ is the dark matter density in units of the
critical density, $\Theta$ gives the ratio of the comoving sound
horizon at decoupling to the angular diameter distance to the
surface of last scattering, $\tau$ is the reionization optical
depth, $n_s$ the scalar spectral index and $A_s$ is the amplitude of
the density fluctuations.

\begin{figure*}[t!]
\resizebox{\hsize}{!}{
\includegraphics[scale=0.8,angle=0]{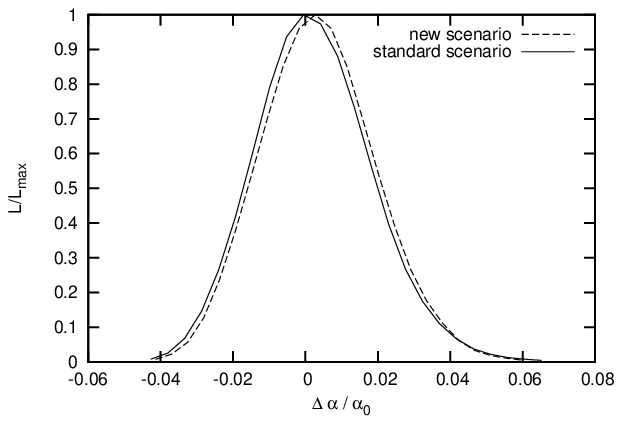}
\includegraphics[scale=0.8,angle=0]{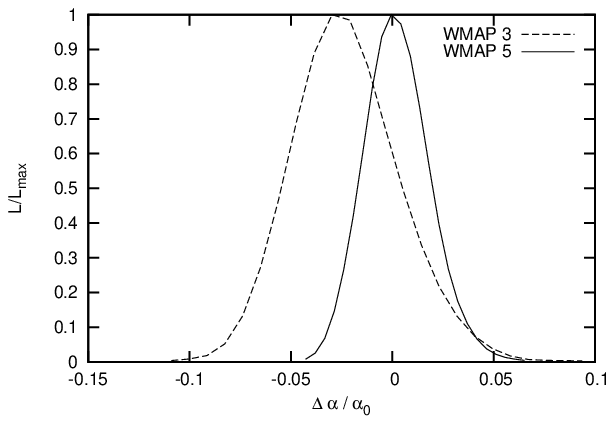}}

\caption{One dimensional likelihood for $\frac{\Delta
\alpha}{\alpha_0}$. Left figure: for WMAP5 data and two different recombination
scenarios.  Right figure: comparison for the standard recombination
scenario, between the WMAP3 and WMAP5 data sets.}
\label{compara_alfa}
\end{figure*}

\begin{figure*}[t!]
\resizebox{\hsize}{!}{
\includegraphics[scale=0.8,angle=0]{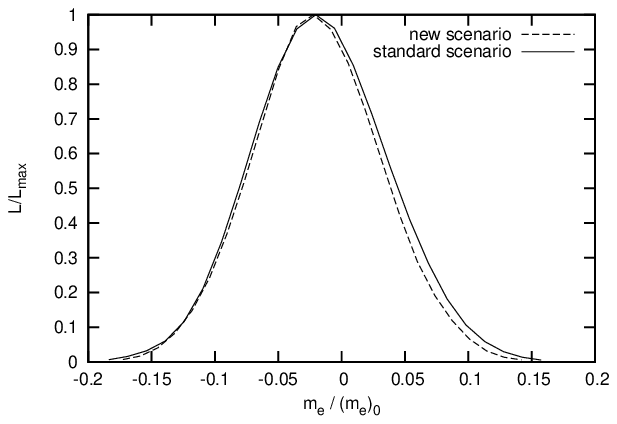}
\includegraphics[scale=0.8,angle=0]{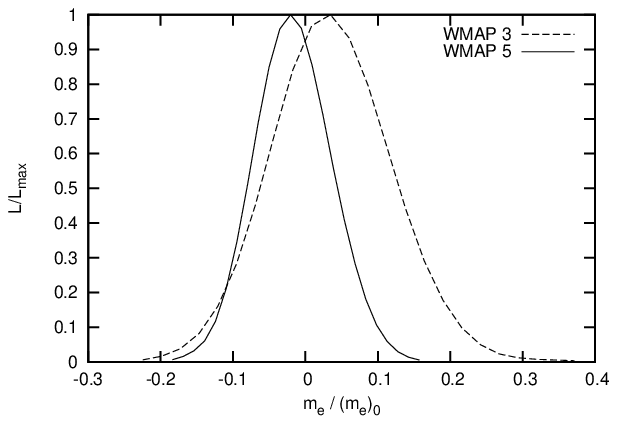}}
\caption{One dimensional likelihood for $\frac{\Delta
m_e}{(m_e)_0}$. Left figure: for WMAP5 data and two different
recombination scenarios.  Right figure: comparison between the WMAP3
and WMAP5 data sets for the standard recombination scenario.}
\label{compara_emasa}
\end{figure*}

In Table~\ref{tablacmb} we show the results of our statistical
analysis, and compare them with the ones we have presented in
\citet{landau08}, which were obtained in the standard recombination
scenario (i.e. the one described in \citep{Seager00}, which we denote
PS), and using WMAP3 \citep{wmap3_temp,wmap3_pol} data.  The
constraints are tighter in the current analysis, which is an
expectable fact since we are working with more accurate data from
WMAP.  The bounds obtained are consistent with null variation, for
both $\alpha$ and $m_e$, but in the present analysis, the $68 \%$
confidence limits on the variation of both constants have changed. In
the case of $\alpha$, the present limit is more consistent with null
variation than the previous one, while in the case of $m_e$ the single
parameters limits have moved toward lower values. To study the origin
of this difference, we perform another statistical analysis, namely
the analysis of the standard recombination scenario (PS) together with
WMAP5 data and the other CMB data sets and the 2dFGRS power
spectrum. The results are also shown in Table~\ref{tablacmb}. We see
that the change in the obtained results is due to the new WMAP data
set, and not to the new recombination scenario. In
Fig.~\ref{compara_alfa} we compare the probability distribution for
$\Delta \alpha/\alpha_0$ in different scenarios and with different
data sets. In Fig.~\ref{compara_emasa}, we do the same for $\Delta
m_e/(m_e)_0$. Bounds on the fundamental constants are shifted to a
region of the parameter space closer to that of null variation in the
case of $\alpha$. On the other hand, limits on the variation of $m_e$
are shifted to negative values, but still consistent with null
variation. The bound on $\Omega_b h^2$ is also shifted to higher
values.

We present here the results of our statistical analysis when only one
fundamental constant is allowed to vary together with a set of
cosmological parameters. We obtained these results using data from
WMAP5, CBI, ACBAR, BOOMERANG, and the $P(k)$ of the 2dFGRS. The
constraints (with 1-$\sigma$ errors) on the variation of $\alpha$ are
$\Delta \alpha/\alpha_0 = -0.002 \pm 0.009$ in the standard
recombination scenario, and $\Delta \alpha/\alpha_0 = -0.001 \pm
0.009$ in the detailed recombination scenario. For $m_e$, both bounds
are $\Delta m_e / (m_e)_0 = - 0.01 \pm 0.03$. The limits are more
stringent than in the case of joint variation of the constants.  This
is to be expected since the parameter space has higher dimension in
the later case. The values for the cosmological parameters are
consistent with those from the joint variation analysis.
 

\begin{table*}
\begin{center}
\renewcommand{\arraystretch}{1.5}
\begin{tabular}{|c|c|c|c|}
\hline
  parameter & wmap5 + NS & wmap5 + PS & wmap3 + PS\\
\hline
 $\Omega_b h^2$ & 0.02241$_{-0.00084}^{+0.00084}$   & 0.02242$_{-0.00085}^{+0.00086}$ & 0.0218$_{-0.0010}^{+0.0010}$ \\
\hline
 $\Omega_{CDM} h^2$ & 0.1070$_{-0.0078}^{+0.0078}$  & 0.1071$_{-0.0080}^{+0.0080}$ & 0.106$_{-0.011}^{+0.011}$ \\
\hline 
$\Theta$ & 1.033$_{-0.023}^{+0.023}$  & 1.03261$_{-0.023}^{+0.024}$ & 1.033$_{-0.029}^{+0.028}$ \\
\hline
$\tau$ & 0.0870$_{-0.0081}^{+0.0073}$   & 0.0863$_{-0.0084}^{+0.0077}$ & 0.090$_{-0.014}^{+0.014}$ \\
\hline
$\Delta \alpha / \alpha_0$ & 0.004$_{-0.015}^{+0.015}$  & 0.003$_{-0.015}^{+0.015}$  & -0.023$_{-0.025}^{+0.025}$\\
\hline 
$\Delta m_e /(m_e)_0$ & -0.019$_{-0.049}^{+0.049}$  & -0.017$_{-0.051}^{+0.051}$ & 0.036$_{-0.078}^{+0.078}$\\
\hline
$n_s$ & 0.962$_{-0.014}^{+0.014}$  & 0.963$_{-0.015}^{+0.015}$  & 0.970$_{-0.019}^{+0.019}$\\
\hline
$A_s$ & 3.053$_{-0.041}^{+0.042}$ & 3.05203$_{-0.04257}^{+0.04269}$  & 3.054$_{-0.073}^{+0.073}$ \\
\hline
$H_0$ &  70.3$_{-5.8}^{+5.9}$  &   70.3$_{-6.0}^{+6.1}$  &  70.4$_{-  6.8}^{+  6.6}$  \\
\hline
\end{tabular}
\end{center}
\caption{Mean values and 1$\sigma$ errors for the parameters
including $\alpha$ and $m_e$ variations. NS stands for the new
recombination scenario, and PS stands for the previous one.} \label{tablacmb}
\end{table*}


\section{Discussion}
\label{conclusion}

The obtained results for the cosmological parameters are in agreement
within $1 \sigma$ with the ones obtained by the WMAP collaboration
\citep{wmap5}, without considering variation of fundamental constants.
It is also interesting to compare our results with the works of
\citet{Nakashima08} and \citet{Menegoni09} where only the variation of
$\alpha$ was analized using WMAP 5 year release and the Hubble Space
Telescope (HST) prior on the $H_0$. In these works, the HST prior on
$H_0$ is used to reduce the large degeneration between $H_0$ and
$\alpha$ and find more stringent constraints on $\alpha$
variation. However, we have shown in \citet{Mosquera07} (where the
variation of $\alpha$ was analysed using the WMAP 3 year release),
that more stringent constraints on $\alpha$ can be found using the
power spectrum of the 2dFGRS. Indeed, our constraints on $\alpha$
alone are more stringent than those reported by
\citet{Nakashima08}. On the other hand, our constraints are of the
same order than those presented by \citet{Menegoni09}, using data from
higher multipoles reported by recent CMB experiments, such as QUAD
\citep{quad} and BICEP \citep{bicep}. Finally, it is important to
stress, that when using the HST prior on $H_0$, the correct value to
be used is the value obtained using only the closest objects, since
bounds obtained using other objects could be affected by a possible
$\alpha$ variation.

\bibliography{bibliografia3}
\bibliographystyle{aa}

\end{document}